# Hybrid Serverless Computing: Opportunities and Challenges[1]

Paul Castro, Vatche Isahagian, Vinod Muthusamy, Aleksander Slominski

IBM Research

## Abstract

In recent years, there has been a surge in the adoption of serverless computing due to the ease of deployment, attractive pay-per-use pricing, and transparent horizontal auto-scaling. At the same time, infrastructure advancements such as the emergence of 5G networks and the explosion of devices connected to Internet known as Internet of Things (IoT), as well as new application requirements that constrain where computation and data can happen, will expand the reach of Cloud computing beyond traditional data centers into Hybrid Cloud. Digital transformation due to the pandemic, which accelerated changes to the workforce and spurred further adoption of AI, is expected to accelerate and the emergent Hybrid Cloud market could potentially expand to over trillion dollars. In the Hybrid Cloud environment, driven by the serverless tenants there will be an increased need to focus on enabling productive work for application builders that are using a distributed platform including public clouds, private clouds, and edge systems. In this chapter we investigate how far serverless computing can be extended to become Hybrid Serverless Computing.

---

[1] **Acknowledgement**: *Preprint of an article submitted for consideration in Edited book (Serverless Computing: Principles and Paradigms) from Springer Publisher.*

**Content**

# 1 Introduction

Cloud computing has evolved from infrastructure services provided by a cloud vendor to a whole array of services needed to develop an application. As more enterprises have migrated their applications to the cloud, there has been a trend towards *hybrid cloud* architectures, where application components are distributed across multiple cloud providers. This could be to deliberately avoid vendor lock-in, due to consolidating applications that were independently developed across heterogeneous platforms, a result of regulatory constraints that necessitate a hybrid on-premise and cloud architecture, or because developers want to take advantage of differentiated features offered by different platforms. As depicted in Figure 1, a natural next step in this evolution is to include edge platforms, IoT equipment, and personal computing devices, an architecture we refer to as *hybrid computing*.

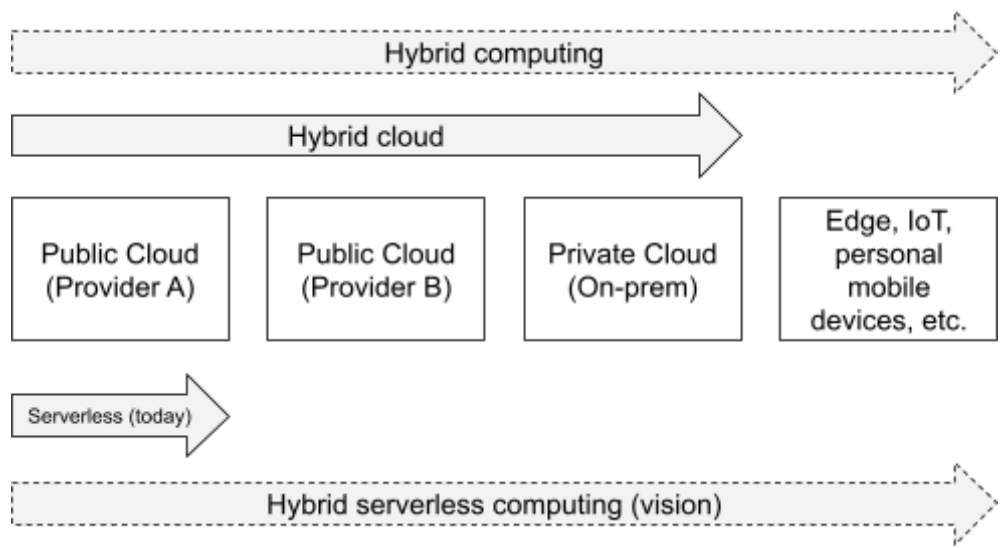

**Figure 1**. Evolution of Hybrid Cloud and related areas

Cloud computing has evolved from running an application on a single public cloud provider to hybrid environments spanning multiple public and private clouds. We see this evolution continuing to encompass devices at the edge and personal computing devices, which we refer to as hybrid computing. Serverless computing today is still largely limited to applications running on a single provider's platform. Our vision is a hybrid serverless computing paradigm where the serverless principles are applied to applications that span multiple clouds and devices.

Serverless platforms arrived less than a decade after public clouds, and can be viewed as a natural extension of cloud but takes a radical step forward to focus on simplicity, scalability, and pay-per-use with little or no knowledge of cloud servers. In serverless [1], most commonly in the form of Functions-as-a-Service (faaS) and newer serverless offerings like Knative Serving [2], users ignore most infrastructure concerns and focus on developing and deploying their application logic.

Unfortunately, serverless applications today are mostly architectured to run on a single vendor's platform. The same technical and non-technical forces steering the evolution towards hybrid cloud and hybrid computing are going to lead serverless application developers to consider more heterogeneous architectures. In our view, there are both business incentives of serverless vendors and technical reasons (discussed in Section 2) that offer headwinds towards a hybrid serverless computing. We chart a course for both serverless practitioners and researchers towards hybrid serverless computing architectures.

We will
- present an overview of the evolution of cloud computing,
- make our case for hybrid serverless computing,
- sketch a path towards hybrid serverless computing,
- and outline challenges in achieving this vision.

In this chapter, we are mostly focused on ensuring developer productivity when building hybrid serverless applications. There are also important runtime performance considerations, which we touch on in Section 5, but are not the core focus of this article.

## 2   Trends in Cloud and Serverless Computing

### 2.1   Cloud and Hybrid Cloud

Cloud evolved from early ideas of Utility [3] and Grid Computing [4]. An early definition of cloud computing has similarities with the definition of serverless computing [5]:

- *The appearance of infinite computing resources available on demand, quickly enough to follow load surges, thereby eliminating the need for cloud computing users to plan far ahead for provisioning.*
- *The elimination of an up-front commitment by cloud users, thereby allowing companies to start small and increase hardware resources only when there is an increase in their needs.*
- *The ability to pay for use of computing resources on a short-term basis as needed (for example, processors by the hour and storage by the day) and release them as needed, thereby rewarding conservation by letting machines and storage go when they are no longer useful"*

That definition of Cloud computing [6] (and others such as [7]) evolved to be "a way of using computers in which data and software are stored or managed on a network of servers". Cloud computing as Infrastructure-as-a-Service (IaaS) began in 2006 with the general availability of Amazon Elastic Compute Cloud in Amazon Web Services (AWS). Today, the term Cloud is often synonymous with commercial public cloud offerings like AWS, Google Cloud, and Microsoft Azure, that offer a catalog of hardware and software that users can provision without any upfront capital costs. Users still have to be aware of, configure, and manage cloud resources, though this can all be done in a self-service manner with no need to manage physical data centers.

The commercial landscape of Cloud Computing has changed significantly since the public availability of AWS in 2006. Today, we are in an era of hyperscale cloud providers, where the Cloud market is dominated by a few planet-scale, public platform providers. Each cloud provider platform is a siloed, vertical stack of software services and infrastructure options designed as general-purpose components to build most cloud native applications. Cloud providers offer a broad catalog of services that are unique to each provider. Services across providers do not cleanly interoperate and today there are few, if any, industry standards. While the cloud providers do provide some managed versions of open source software, the tooling and features of these will differ between providers. Skills gained by using one provider do not necessarily translate to another.

In practice, this lack of interoperability may dictate the need to move beyond a few hyperscale cloud providers. Many enterprises make use of more than one cloud provider [8][9]. This could be deliberate; users will naturally want to build multi-cloud or hybrid cloud solutions that avoid vendor lock-in and take advantage of the best services available, or just a result of the organic, uncoordinated nature of cloud adoption in some industries [10]. Larger enterprises have significant capital investments in their own private data centers and in many cases it makes sense to adopt a hybrid mixture of on-premise and public clouds. Also, increasing awareness of data privacy, industry specific requirements, and evolving regulatory issues are now a pressing concern - efforts like GAIA-X [11] seek to restrict the use of data and computation in order to preserve data sovereignty. As such, the hyperscale cloud providers and other 3rd party vendors are looking to ease the burden of creating multi-cloud or hybrid cloud solutions.

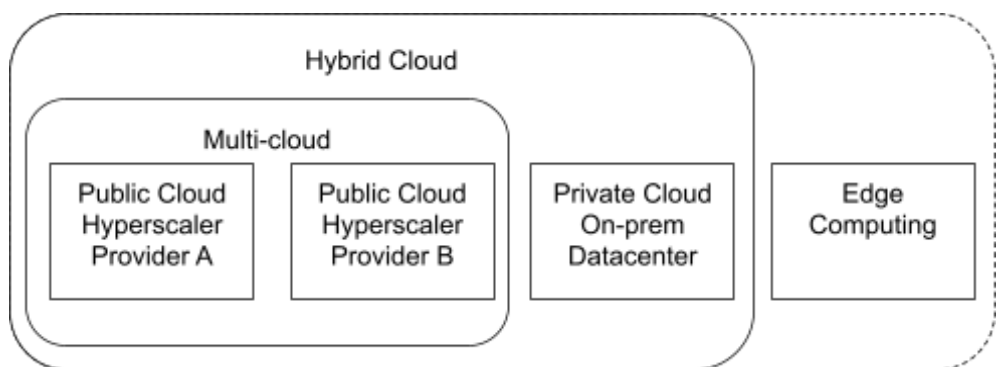

**Figure 2**. Relations between Public Cloud, Private Cloud, Multi-Cloud and Hybrid Cloud

There are many terms - multi-cloud, hybrid cloud, polycloud, supercloud - being defined that capture different perspectives on this emerging, heterogeneous, distributed platform [12]. In this chapter, we refer to Hybrid Cloud in the broadest sense, encompassing public clouds, on-premise data centers, and edge services. At IBM, hybrid cloud "combines and unifies public cloud, private cloud and on-premises infrastructure to create a single, flexible, cost-optimal IT infrastructure" [13]. Microsoft has a similar definition [14]: "A hybrid cloud—sometimes called a cloud hybrid—is a computing environment that combines an on-premises datacenter (also called a private cloud) with a public cloud, allowing data and applications to be shared between them. Some people define hybrid cloud to include "multi-cloud" configurations where an organization uses more than one public cloud in addition to their on-premises datacenter.

Many in academia and industry are taking the perspective of Hybrid Cloud as a distributed application development platform not tied to a single provider with applications composed out of heterogeneous services running anywhere. For example, the vision of Sky Computing [15] is inspired by the evolution of the Internet and the goal would be to connect the offerings of cloud providers and other heterogeneous cloud platforms through the use of industry-wide protocols. The RESTless Cloud [16] invokes the need for an industry-wide POSIX-like standard to cloud APIs to ease interoperability. Similarly, User Defined Cloud [17] looks for high-level declarative techniques for users to define their own clouds based on application requirements. In industry for example Dell Technologies APEX [18] and HPE GreenLake [19] market hybrid cloud solutions. An open source Crossplane [20] project allows users to control resources in different clouds using a high-level interface running in Kubernetes.

Even though the definitions of Hybrid Cloud are evolving and there are multiple ideas for its future we can see that this area is gaining in popularity alongside serverless computing. By looking at Google trends [21] we can see the growth in popularity of "Serverless computing" and "Hybrid Cloud" as search terms in Figure 3.

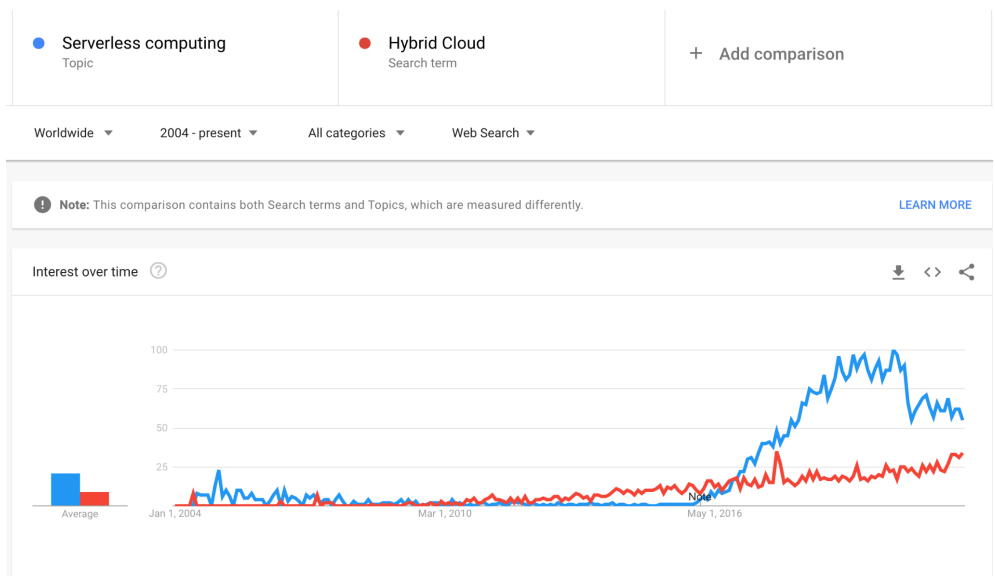

**Figure 3**. Popularity of "Serverless computing" and "Hybrid Cloud" in the last 15 years.

## 2.2 Computing trends toward Hybrid Computing

It is estimated that there are between 10-100 milion servers used in all of the public clouds [22] and many more in private clouds and on-prem data centers. There are about 1-3 billion personal computing devices used worldwide [23][24].

Modern devices are connected and support more and more advanced computing capabilities. There is a large amount of computation resources available today in hybrid clouds and servers on prem but many more computing devices are not servers but personal computing devices (PCs) such as laptops, with the number of smartphones and IoT devices growing very quickly . According to current research [25] "In 2022, the number of smartphone users in the world today is 6.648 Billion, which translates to 83.96% of the world's population owning a smartphone. In total, the number of people that own a smart and feature phone is 7.26 Billion, making up 91.69% of the world's population." The number of IoT devices already is several times more than the number of people and growing very fast: in 2021 there were more than 35 billions of IoT devices and double of this number is predicted by 2025 [26].

Important part of connecting the cloud with consumer and IoT devices is edge computing providing low-latency high-bandwidth intermediate computing capabilities. Edge computing is also growing fast, according to IDC [27]: "the worldwide edge computing market will reach $250.6 billion in 2024 with a compound annual growth rate (CAGR) of 12.5% over the 2019–2024 forecast period.

The computing devices are getting cheaper, more energy efficient, and almost always connected to the Internet. We summarize the number of different types of computing devices in Table 1.

| Computing Paradigm | Number of Devices (in billions) |
| --- | --- |
| Hybrid Cloud | 0.1-1B |
| Personal Computing (PC) | 1-3B |
| Smartphones | 6-7B |
| IoT | 35B |

**Table 1**. Number of computing devices in 2022

To describe all those different types of computing we introduce the term **Hybrid Computing**. It is a superset of Hybrid Cloud to include not only cloud or edge computing but any kind of computing that is connected in some ways to the Internet (may be not persistent connection). Additional types of computing may become more important or emerge in future such as Fog Computing, Web3 and Blockchain based distributed computing, metaverse with AR/VR devices, and so on. In Figure HC we show how Hybrid Computing is related to Hybrid Cloud and different types of computing paradigms.

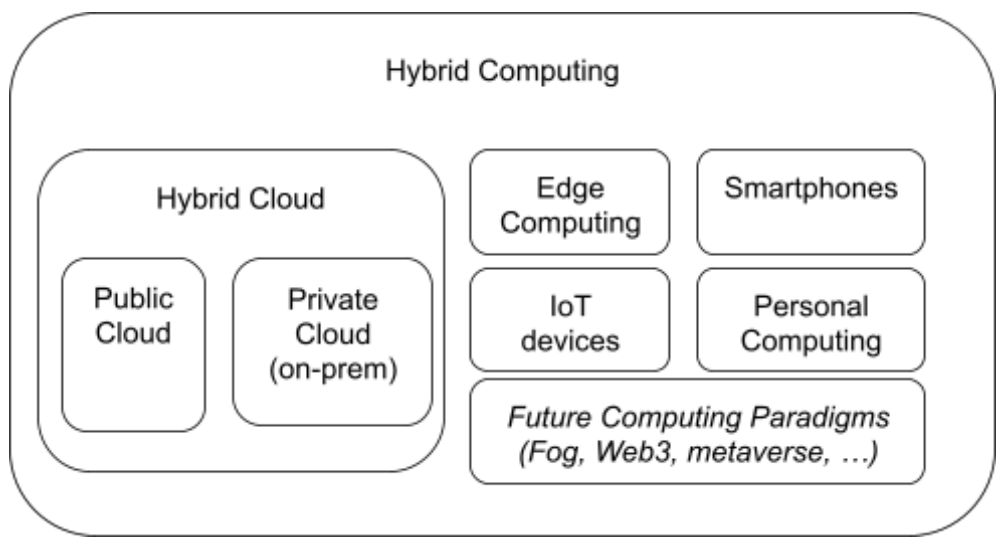

**Figure 4**. Hybrid Computing as a superset of Hybrid Cloud

**Definition: Hybrid Computing combines any computing paradigm to allow creation of applications that can operate in a seamless way in multiple computing infrastructures connected by the Internet.**

It is our hypothesis that whatever future computing paradigms may be they are going to need serverless characteristics: simplicity, easy scalability, developer focus etc.

## 3     The case for Hybrid Serverless Computing

We define serverless *computing as a platform that hides server usage from developers and runs code on-demand automatically scaled and billed only for the time the code is running* [28]. This broad view of serverless computing emphasizes the benefits of hiding complexity and offloading much of the work around application scaling, security, compliance, and other infrastructure management concerns to the cloud provider. To paraphrase Tim Wagner, the former GM for AWS Lambda and considered by some as the father of serverless compute, emphasized that serverless greatly enhances productivity because it "gets rid of undifferentiated heavy lifting" for application developers and allows for "capital efficient value creation" [29]. Recent work [30] predicts that serverless will be the dominant form of cloud computing and that the "future evolution of serverless computing, and in our view of cloud computing, will be guided by efforts to provide abstractions that simplify cloud programming." With this broader interpretation of serverless and the economics associated with serverless, one can argue that it represents an evolution of the cloud platform itself and is a key element to attract the next generation of cloud native developers to the cloud.

In our view, serverless' focus on reducing complexity and allowing the transparent and elastic use of compute resources provides an ideal development platform that federates the heterogeneous, distributed hybrid cloud platform,

While much has been made of the pay-per-use and transparent auto-scaling properties of serverless compute platforms, the largest benefit to developer productivity may be raising the level of abstraction for developers so their interface to the cloud is programmatic and requires minimal configuration to conform to system-defined limits. Serverless computing, however, is still in its infancy, with most serverless applications developed for and run on a single serverless platform. There is a class of enterprise applications that aren't amenable to run fully on a public cloud due to regulatory constraints [31], inconsistent development experience across vendors [32], and the vendor and platform lock-in in today's most popular serverless platforms is holding back developers [33][34].

Current serverless platforms including AWS Lambda, Azure Functions, Google Cloud Functions, IBM Cloud Functions, and offerings based on open source software like OpenFaaS, Apache OpenWhisk, and Knative, are managed services and do not interoperate across cloud provider boundaries. This is reflected by the current use of serverless applications which typically use a single platform because:

1. Serverless relies heavily on a cloud provider's integrated ecosystem of databases, storage, etc.. which is mainly not serverless.
2. Serverless applications are mostly built as extensions of already deployed cloud applications, to compose cloud services, or as features of cloud applications.
3. Its relatively easy to end up in anti-patterns of serverless applications design [35]

There is little or no expectation of interoperability across different serverless platforms. This has forced developers who already lack the tools to deal with continuous changes of the cloud into the need for an accelerated set of skills to build and debug serverless functions. Developers that want to build serverless applications, or any application for that matter, on a hybrid cloud platform have little help. Current multi-cloud management tools focus more on operational tasks around managing multiple Kubernetes clusters. For example SUSE Rancher [36] helps manage multiple kubernetes clusters, while the CNCF incubating project Crossplane [20] focuses on cross-provider resource management through Kubernetes operators.

One exception is Funqy [37], a Java-based library that allows you to write Java functions using a single API that works across FaaS implementations using the Quarkus runtime [38]. While this potentially improves code portability, it does not address the broader issues of hybrid cloud development.

## 3.1 Benefits to application developers and platform providers

A hybrid serverless architecture brings a number of benefits both to the serverless application developers and platform providers.

- Developers benefit from the flexibility to migrate their serverless applications to different platform vendors to take advantage of pricing differences. The choice of vendor is not a one-time decision, since an application may exhibit different load patterns as it evolves and different vendors may offer pricing plans optimized for different classes of workloads.
- Similarly, an application may require different non-functional guarantees (such as throughput targets) or require different platform services (such as a global load balancer) as it matures from a proof-of-concept to a popular production application. Developers need flexibility to migrate vendors as their requirements evolve.
- A related point is that developers may want to deploy portions of their applications to the edge to improve user-perceived latency, Or leverage a combination of VM-based and serverless based resources for application execution [39]. A hybrid architecture

- would facilitate more fine-grained deployment choices for portions of the application.
- Similarly, developers may have constraints on where parts of their application can be deployed. For example, there may be regulations that govern where health records can be stored, whether within a country or an on-premise datacenter managed by the company.
- Another developer benefit is the ability to mix and match services provided by different platform vendors rather than make the choice to deploy their entire application on a single platform. For example, they may choose the FaaS platform of one vendor, the logging service from another vendor, and the messaging service from a third one.

Regarding the last point, vendors may discourage such hybrid architectures with their pricing plans, and there may be legitimate performance reasons to deploy an entire application on a single platform, but we still think that developers should have choice in their deployment options. Furthermore, we believe that giving developers this choice can in turn benefit platform vendors.

In particular, serverless platforms today need to offer a breadth of services that developers may need. Making hybrid architectures more developer friendly gives an opportunity for smaller vendors to offer specialized services without having to replicate the complete stack of services needed for an enterprise serverless application. For example, nimble vendors may offer quantum compute-based FaaS service or predictive complex event processing service that developers can easily integrate into their overall serverless application.

Developer flexibility in this case can lead to more competition among platform vendors, resulting in a virtuous cycle of lower costs, more innovative service offerings, and increased adoption of hybrid serverless architectures.

## 3.2    Roadblocks to hybrid serverless computing

Borrowing the hourglass metaphor from computer networking, let's consider the waist of the hourglass to be the interface that the cloud vendor offers to application developers. As illustrated in Figure 5, in an era when developers were building applications using VM abstractions, the waist was relatively low in the cloud application stack. In the serverless era, however, the vendor takes on many more responsibilities under the serverless interface (such as FaaS).

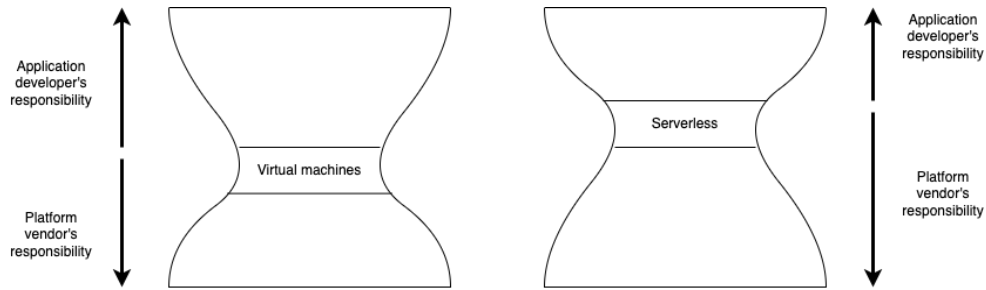

**Figure 5**. Building cloud applications over a virtual machine abstraction requires the developer to be responsible for more of the application stack. Conversely, serverless abstractions give the platform vendor opportunities to optimize more of the stack.

The fact that the vendor owns a much larger part of the application stack leads to a natural tendency for single-vendor serverless applications. There are technical reasons for this, such as the vendor's visibility into the larger stack affording them the ability to implement platform-level optimizations. There are also business incentives, such as pay-as-you-go the serverless billing models only becoming economical if the vendor has the ability to monitor and manage the underlying resources.

The challenge is how to break out of this ingrained bias towards single-vendor serverless applications. In this chapter, we argue that there are two approaches to building hybrid serverless applications that span the services offered by vendors. The compiler-based approach does not require any special support from vendors, whereas the standards-based one is more robust but requires vendor support.

## 4   Towards a Hybrid Serverless Architecture

It is possible today to use multiple clouds and multiple serverless offerings (sometimes called hybrid serverless multi-cloud [40]) but it is not easy. The main limitations are due to lack of standards and shared abstractions.   Recent research work considers Hybrid Serverless deployments; for example [41] discusses the merits of a virtual serverless provider to aggregate serverless platforms across clouds. [42] federates endpoints to execute serverless functions in the domain of scientific computing. [43] goes a step further to enable not just the deployment of functions on distributed endpoints, but also scale those endpoints automatically.

Examples of AWS RedShift and emergence of Snowflake [44] validate the need for solutions that run across multiple cloud environments [45].

Recent proposals emphasize consistent developer experience across multiple clouds either as "supercloud" [46] or Sky computing [15]: "the vision of a Sky full of computation, rather than isolated clouds .

We believe that a combination of evolving computing trends, hybrid cloud approaches and serverless computing may lead to one common computing approach: **Hybrid Serverless Computing**.

**Definition: Hybrid Serverless Computing extends Hybrid Cloud to include all computing paradigms unified by use of containers and providing serverless approach to computing by using standards or compiler-based approach.**

We will discuss a possible way to build hybrid serverless computing, what we have today and what is missing.

Observation: Most serverless applications today primarily use the services of a single vendor. This is because of the ease-of-use and performance benefits of using a coherent ecosystem of services in a vendor's platform.

Hypothesis: There won't be a single vendor that will provide all the services with the desired functional and non-functional properties that an application developer wants. There is too much ongoing innovation in cloud services to expect that a single vendor will offer the best-of-breed service along all dimensions.

Assertion: There will be a need for both standards-based and compiler-based approaches.

## 4.1 Two approaches to achieving Hybrid Serverless Computing: Compilers and Standards

### 4.1.1 Running example

Consider a simple "Hello, World" serverless application that consists of a function to process images. The function may generate a thumbnail of the image or convert it to another format. In our example, we consider a function to automatically enhance an image, such as adjusting exposure or changing the sharpening settings. Multiple cloud-based photo organization services provide such a feature.

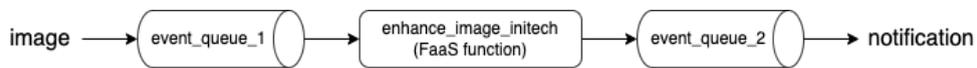

**Figure 6**. A simple serverless application running on Initech's platform.

In the figure above, we see that the "enhance_image" serverless function is triggered by new images arriving in an event queue. The output enhanced images are emitted to another event queue, which may trigger other functions to store the image or notify the user of the new image.

This architecture assumes that all the components, including the queues and functions are deployed on a single vendor's serverless platform. Let's call this vendor Initech. Consider the case where another vendor, Acme, offers a more desirable image enhancement service. It may be that Acme's service uses a more advanced image processing AI model, is cheaper to operate, or runs on special hardware to offer lower latency. The following sections will consider increasingly more sophisticated ways in which to incorporate the Acme service into this architecture.

### 4.1.2 Implementation 1: Blocking API call

In this implementation, we add a function to the application deployed on Initech's platform to make a blocking call to Acme's image enhancement service.

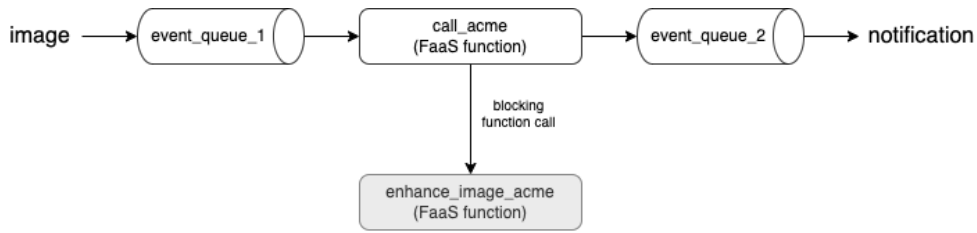

**Figure 7**. A hybrid serverless application that spans two vendor platforms. The clear elements are those deployed on Initech's platform, while the shaded elements are those running on Acme's platform.

The advantage of this implementation is that it is a relatively straightforward change that does not require modifications to any other part of the application, nor any special requirements of Initech's platform or Acme's service.

The disadvantage is that this is not a pure serverless application. In particular the blocking function call is an anti-pattern that results in wasted compute and double-billing [47].

### 4.1.3   Implementation 2: Event bridge

In this next implementation, Acme's service is now more serverless-friendly. We assume that neither vendor allows direct access to the queues in their platform from outside, but do allow FaaS functions to be triggered from external calls.

The resulting architecture, shown in Figure 8, includes a number of additional queues and functions in both vendors' platforms. In particular, there is an event queue in Acme's platform that can trigger a serverless function to perform the image enhancement. And the function output is pushed to an event queue that triggers a function to send the enhanced image back to the rest of the application pipeline in Initech's platform.

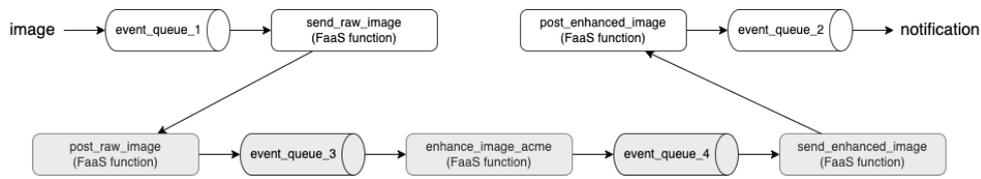

**Figure 8**. A hybrid serverless application that adheres to serverless best-practices.

The advantage of this implementation is that it is now a pure serverless architecture, taking advantage of event-driven design patterns.

The disadvantage, which is apparently comparing Figures 7 and 8, is the additional architectural components added to bridge the event queues between Initech and Acme's platforms. These are not part of the core application logic.

Another disadvantage is that the additional components are only there in the case of the hybrid platform. If we revert back to using Initech's image enhancement services, these additional components are superfluous. Adding and removing these components depending on which service the developer chooses to use adds to the deployment complexity.

### 4.1.4   Implementation 3: Event standard

In this implementation, we suppose that eventing is a common enough service that there is an eventing standard that all serverless platform providers adhere to. Among other things, this means that Acme's serverless function can subscribe to and be triggered by events in Initech's event queue, and Acme's serverless function can publish events to Initech's event queue.

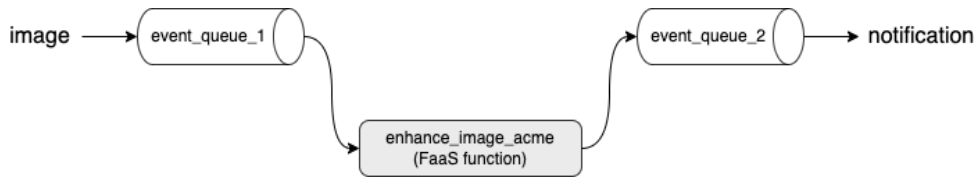

**Figure 9**. Having vendors support a common eventing standard simplifies the hybrid serverless architecture.

The advantage of this implementation is that it results in a far simpler architecture, with no extra components that arise from deciding to deploy the application across multiple vendor platforms.

The obvious disadvantage of this approach is that vendors need to adopt and adhere to an eventing standard.

An even more fundamental challenge is that an eventing standard needs to be defined. This may be an existing event interface, such as Knative Eventing [48], that becomes an ad-hoc standard, or an official standard published by a standards body. The interface and semantics of a serverless eventing standard that supports the range of environments we want to support – including resource constrained IoT devices, secure on-prem data centers, and powerful public clouds – remains an open research challenge.

### 4.1.5 Implementation 4: Event bridge with Compiler

In this implementation, a compiler is used to generate the glue components to wire the application logic between Initech and Acme's platforms.

Notably, as depicted in Figure 10, the developer's application logic should look like the one in the Event Standard implementation, while the deployed architecture output by the compiler looks like the one in the Event Bridge implementation.

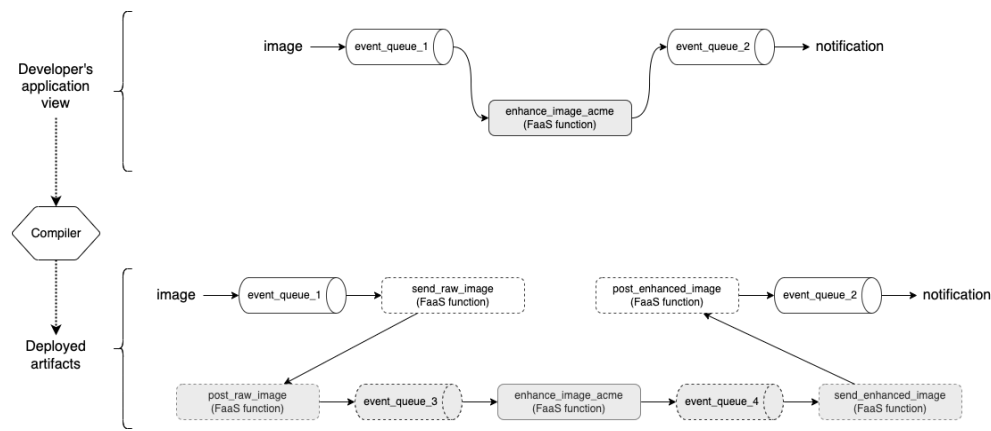

**Figure 10**. In lieu of a common eventing standard, a compiler can help generate the artifacts needed to bridge vendor platforms. The generated artifacts, which are not part of the core application logic, are depicted with a dotted outline.

The advantage of this implementation is that it hides from the developer the architectural complexities that are not related to the application logic.

The disadvantage is that it requires someone to build the compiler to perform this task. Furthermore, the compiler needs to have native understanding of the capabilities of each vendor platform, and hence developers are limited to those vendors supported by the compiler.

### 4.1.6   Implementation 5: Offline service selection

In this implementation, we now assume that image enhancement is a common enough service that there is a well-defined serverless-friendly standard for this service that vendors adhere to. In other words, this service becomes a first class capability in a serverless platform. Note that in practice there may be other more core backend services, such a key-value storage service, that are likely to become a first class standard before something like an image enhancement service.

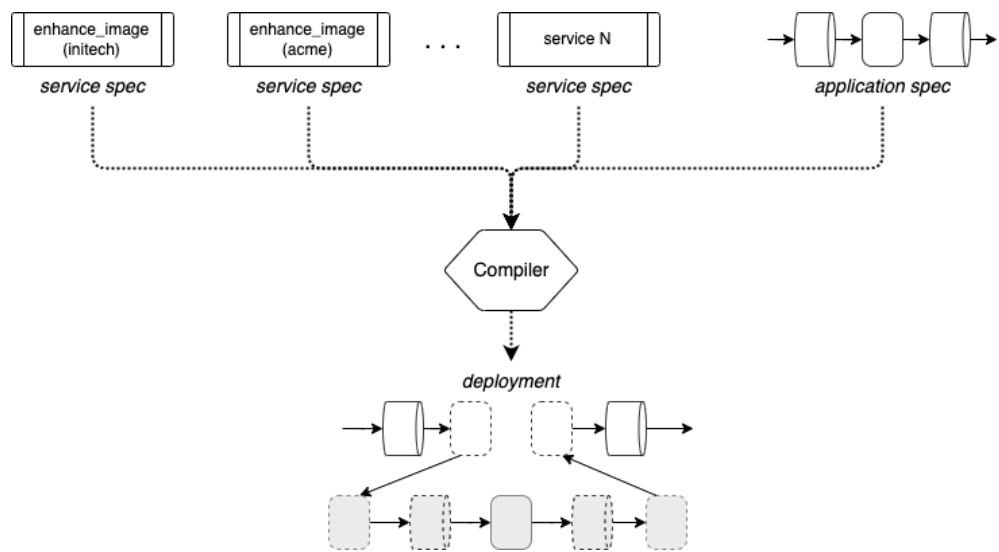

**Figure 11**. The presence of serverless-friendly standards for service interface specifications and application requirements allows the compiler to select the optimal service across platforms.

The new capability that this implementation offers is the ability to consult a catalog of functionally equivalent services and automatically select the optimal one according to the non-functional requirements of the application. For example, suppose Initech's version of the image enhancement service is slower but cheaper than Acme's. If the developer knows that offline batch processing is acceptable for the application's use case, they can specify a latency requirement in the order of hours, and let the compiler choose the cheaper Initech service. On the other hand, an application with low latency requirements would choose to use Acme's service. In either case, the vendor-specific artifacts are automatically generated and deployed.

The ability to perform service selection requires a standard to describe the services, including functional specifications such as the input and output parameters, and non-functional ones such as the cost per invocation and response time estimates. While there have been service interface standards defined in the past, such as WSDL [49] or recently proposed AsyncAPI [50], they are not in common use today. We believe there is still an open question on what service interface specification should look like, especially serverless-specific concerns such as cold-start latencies.

There is also a requirement for a standard way for application developers to define the requirements of their application. This too is an open research challenge and needs to include aspects such as the pricing or latency constraints.

Once there are standards for the service interfaces and applications, a compiler can make deployment and service selection decisions. For example, suppose Initech's version of the image enhancement service is slower but cheaper than Acme's. If the developer knows that offline batch processing is acceptable for the application's use case, they can specify a latency requirement in the order of hours, and let the compiler choose the cheaper Initech service.

The design of the compiler's service selection algorithm is also an open research challenge. It is not clear how the techniques in the literature [51] need to be extended to consider serverless-specific architectural requirements, such as the overhead of additional queues and functions when crossing vendor boundaries, and the span of hybrid environments, such as possible intermittent connectivity when interacting with IoT devices.

#### 4.1.7 Implementation 6: Runtime optimizer

This implementation is an extension of the one above, but has a component that monitors an application at runtime and makes changes to the deployment to maintain the application requirements.

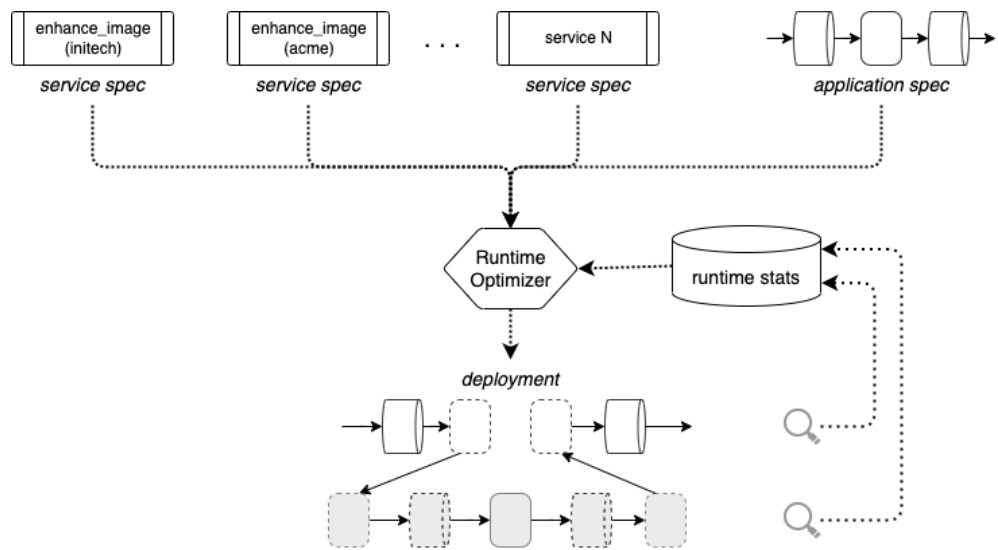

**Figure 12**. An optimizer monitors application statistics at runtime and makes redeployment decisions based on changing service profiles and requirements.

In addition to the service and application specs required in the Offline Service Selection implementation above, we now need a standard way to instrument each vendor's platform to monitor the application behavior at runtime.

In addition, this solution requires a standard way to deploy and configure services in a vendor's platform. This can be optional, since there is an alternative where vendor-specific support is added to the compiler. This latter approach does not change the application developer's user experience but limits the set of vendors that the runtime optimizer can work with.

There is an open research question on the design of the optimizer along at least two dimensions. First, as with the previous implementation, the optimization algorithms can borrow from prior art [mape-iot], but need to be extended for serverless-specific behaviors and hybrid platform properties. Second, the optimizer is a runtime component and should itself be architected with serverless best-practices in mind.

## 4.2 Discussion

The sequence of implementations for a hybrid serverless application above required a combination of custom implementations by the developer, support from a compiler, and the presence of standards that vendors adopt. The table 2 below summarizes these capabilities. The choice of implementation will be based on the maturity of standards available and adopted by vendors.

| Implementation | Custom code | Compiler support | Required standards |
|---|---|---|---|
| 1: Blocking API call | API call | | |
| 2: Event bridge | Eventing | | |
| 3: Event standard | | | Eventing |
| 4: Event bridge with Compiler | | Eventing | |
| 5: Offline service selection | | Service selection | + Service, application |
| 6: Runtime optimizer | | Online replanning | + Monitoring, deployment |

**Table 2**. A summary of the capabilities that were implemented manually or with the help of compilers and standards.

We believe that new capabilities will typically follow a pipeline, first being implemented manually, such as the eventing bridging architecture in Implementation 2. As design patterns form around common architectural designs, they will get adopted by compilers, and as these patterns mature, the community will settle on ad-hoc standards or define official ones to support these architectural patterns.

The table 3 below illustrates this progression of features. Notice that not all features, such as feature_4, may never get promoted to compiler support, and others, such as feature_2, may never get adopted as standards. This progression is by no means unique to hybrid serverless applications, but provides a useful framework to drive towards the hybrid serverless computing vision despite community inertia or vendor opposition.

|        | Implemented manually   | Supported by compilers   | Adopted as standards   |
|--------|------------------------|--------------------------|------------------------|
| Time 1 | feature_1              |                          |                        |
| Time 2 | feature_2              | feature_1                |                        |
| Time 3 | feature_3              | feature_2                | feature_1              |
| Time 4 | feature_4              | feature_2, feature_3     | feature_1              |
| Time 5 | feature_4, feature_5   | feature_2                | feature_1, feature_3   |
| Time 6 | feature_4              | feature_2, feature_5     | feature_1, feature_3   |

**Table 3**. An illustration of how capabilities exercised by hybrid serverless application architectures will get promoted from manually being implemented by application developers, to being supported by compilers, and finally getting adopted as ad-hoc or official standards.

Where standards don't exist, the compiler-based approaches will help application developers build hybrid applications without having to bother with the particulars of bridging across vendor platforms. Implementation 4 is an example of this approach.

When standards are more mature, we can consider other approaches. For example, having vendors agree on only an eventing standard leads to Implementation 3, whereas a full set of standards on the services and platform supports the full hybrid serverless vision in Implementation 6.

We believe that the landscape of serverless platforms is evolving too quickly now to expect standards to cover all the functional and non-functional specifications of services and applications requirements.

For example, vendors may want to offer platforms that are HIPPA-certified, run on green-energy, offer specialized hardware for ML training workloads, or provide quantum compute instances. All these new innovations will be supported first by manual implementations by developers, then by the compiler-based approach where compilers have baked-in knowledge of these vendor-specific capabilities. As these innovations commoditize, standards will emerge and developers can transition to the standards-based approach. There will be an ongoing transition from compiler-based to standards-based.

## 5    Opportunities and Challenges

A hybrid serverless model brings with it a number of challenges across the stack that span across multiple disciplines such as data placement and migration issues. Furthermore, there is a need to address the impedance mismatch when bridging across serverless platforms from multiple providers, including the non-functional properties such as latency, scalability, availability, and cost. For example, While cold-start behavior has been studied within the context of serverless functions deployed on a single platform [52] [53], it is not clear what is the emergent cold-start behavior when a serverless function running on one platform calls a function on another. There are also functional mismatches, such as security policies, and messaging semantics that need to be reconciled. In this section, we will describe challenges and outline open research problems.

### 5.1    Hybrid Serverless Computing Research Challenges in Academia and Industry

We think there are several opportunities for research in academia and industry that may be critical to evolving serverless computing and hybrid cloud (ordered by complexity and size).

**Emerging standards**: to what degree there can be standards used by hybrid serverless computing? For example can Cloud Events [54] be used with AsyncAPI [50] to support hybrid serverless computing?

**Interfaces description**: what the interface to custom services should look like. It should include functional specifications, such as the input and output parameters, and non-functional ones, such as the cost per invocation and response time estimates. What is common across multiple services and vendors? What could be standardized?

**Application and solutions descriptions**: research ways for application developers to define the requirements of their application. What is common across vendors and could be standardized for vendor runtime monitoring and vendor deployment interfaces? Compare existing and propose new approaches.

**Container adaptation**: can containers be adapted to run in all possible computing environments, from public cloud to edge to IoT devices? How to manage tradeoffs in resources (CPU, memory, network) vs cost and energy usage. Can security be maintained across multiple environments in one simple consistent way? Can new upcoming packaging formats like WebAssembly [55] be used?

**Container orchestration**: cold-start and other challenges (such as security when compared to virtual machines) must be addressed to make container-based and container orchestration successful long-term

**Compiler optimizations**: how the compiler optimizes the selection of services based on the application requirements. What would be the designs and implementations of such compilers? What optimizations could be applied? How transparent such compiling and optimizations are for developers and to what degree they can support observability and developer productivity? For example could monitoring be done in a standard way and still support multiple platforms?

**Tooling, Productivity and Developer Experience**: Cost of running containers with 1 GB of memory per hour will be less than $0.01 in the near future (about $0.06 in 2022) [56]. However, one cost is unlikely to change which is developer time per hour. Therefore it is crucial for hybrid serverless computing to support consistent developer experience regardless where containers are running: local, remote, cloud or edge should be running identical code and use the same same tooling that supports use of standards, consistent local and remote experience (aka transparency [57]), observability, debuggability, etc. There will be differences but common ground based on standards is the only way to have a success similar in scope to the Internet.

# 6 Conclusion

There exists a huge opportunity to build one open-standards and open-source based hybrid serverless computing platform that will have as large an impact as the Internet by providing computing standards to allow building hybrid serverless computing solutions anywhere. Parts of hybrid serverless computing may not be based on open standards when there are computing niches dominated by existing providers. Still hybrid serverless computing may provide one set of abstractions and standards that provide alternatives that may win in the longer term. It is unlikely that one provider can build hybrid serverless computing solutions instead it is most likely that the best solutions will require a diverse set of computing resources and will need standardization to support hybrid serverless computing.

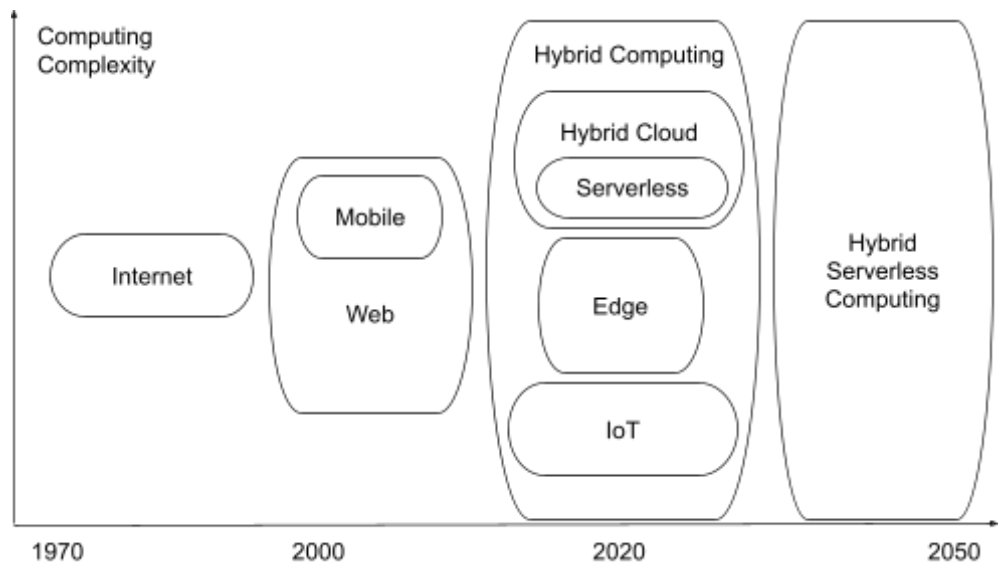

**Figure 13**. An optimizer monitors application statistics at runtime and makes redeployment decisions based on changing service profiles and requirements.

We expect that over a longer period of time different types of computing (Serverless, Cloud, IoT, etc.) will become more and more similar eventually creating a Hybrid Computing paradigm (see Figure 13) that will be as easy to use as Serverless Computing is today leading to Hybrid Serverless Computing.